\documentstyle[emulateapj,psfig]{article}

\makeatletter
\newcommand{\chan}{{\sl Chandra}}
\newcommand{\xmm}{{\sl XMM-Newton}}
\def\ls{\lower 2pt \hbox{$\;\scriptscriptstyle \buildrel<\over\sim\;$}}
\def\gs{\lower 2pt \hbox{$\;\scriptscriptstyle \buildrel>\over\sim\;$}}

\slugcomment{The Astrophysical Journal, v.643, 2006 June 1, in press}
\begin{document}

\title{The Pulsar Wind Nebula of the Geminga Pulsar}

\author{G. G. Pavlov\altaffilmark{1}, D. Sanwal\altaffilmark{1}, and V. E. Zavlin\altaffilmark{2}}
\altaffiltext{1}{Pennsylvania State University,
525 Davey Lab., University Park,
PA 16802; pavlov@astro.psu.edu}
\altaffiltext{2}{Space Science Laboratory, NASA MSFC SD59, Huntsville, AL 35805}

\begin{abstract}

The superb spatial resolution of {\sl Chandra} has allowed  us to detect 
a $20''$-long
tail behind the Geminga pulsar, with a hard spectrum (photon index
$\Gamma = 1.0\pm 0.2$) and a luminosity
$(1.3\pm 0.2)\times 10^{29}$ ergs s$^{-1}$ in the 0.5--8 keV band, for an assumed
distance of 200 pc. 
The tail could be either a pulsar jet, 
confined by a toroidal magnetic field of $\sim 100\,\mu$G, or it can be
associated with the shocked relativistic wind behind the supersonically moving
pulsar confined by the ram pressure of the oncoming interstellar medium.
 We also detected an arc-like structure $5''$--$7''$ ahead of the
pulsar, extended perpendicular to the tail, with a factor of 3 lower luminosity.
We see a $3\,\sigma$ enhancement in the {\sl Chandra} image apparently connecting
the arc with the southern outer tail
that has been possibly detected with \xmm.
The observed structures imply that the Geminga's pulsar wind
is intrinsically anisotropic.
\end{abstract}


\section{Introduction}

It has been recognized long ago that pulsars steadily lose their
rotational energy via relativistic pulsar winds (PWs).
The PW shocks in the ambient medium and forms a pulsar-wind
nebula (PWN) which emits synchrotron radiation
(Rees \& Gunn 1974).  
An isotropic outflow from a stationary pulsar in a uniform medium forms a
spherical termination shock (TS) at a radius $R_s\simeq (\dot{E}/4
\pi c p_{\rm amb})^{1/2}$, where $\dot{E}$ is the pulsar's spin-down power and
  $p_{\rm amb}$ is the ambient pressure.  For a pulsar moving at a speed $v$
with respect to the medium, the TS shape is distorted by the ram pressure,
$p_{\rm ram} = \rho_{\rm amb} v^2$.  At supersonic speeds, when the ram pressure
exceeds the ambient pressure, the TS of an isotropic PW acquires a bullet-like
shape (Bucciantini,
Amato, \& Del Zanna 2005 [hereafter B05], 
and references
therein) with a distance $R_{\rm h} \simeq (\dot{E}/4\pi c p_{\rm ram})^{1/2}$
between the pulsar and the bullet head.  The shocked PW is confined between the
TS and the contact discontinuity (CD) surface which separates the shocked PW from
the shocked ambient medium between the CD and the forward bow shock (FBS).
The shape of the shocks and the overall appearance of the PWN depend on the
interplay of $\dot{E}$, $p_{\rm amb}$, $\rho_{\rm amb}$, and $v$.  Thus, PWNe
produced by fast-moving pulsars provide an important 
diagnostic tool for studying PWs,
the ambient medium, and pulsar velocities.

X-ray PWNe have been observed around about 30 pulsars.  High-resolution
observations with {\sl Chandra} show that the PWN structure is never a simple
spherical shell.  In particular, young PWNe (e.g., the Crab and Vela
PWNe;
Weisskopf et al.\ 2000; Pavlov et al.\ 2003) are often approximately axially
symmetric, with jets along the symmetry axis (which apparently coincides with
the pulsar's spin axis)
and a torus-like structure around
the axis (an equatorial PW outflow).  There are some X-ray PWNe whose cometary
shape is clearly caused by the pulsar motion. Some of them (e.g., the Mouse PWN,
powered by the young PSR J1747$-$2958; Gaensler et al.\ 2004) are confined
within a bow-like boundary, but they do not show a shell structure, in contrast
to the sharp H$_\alpha$ ``bows'' emitted by the shocked ambient gas at FBSs
produced by some pulsars.  {\sl Chandra} observations of the pulsars B1757--24
(Kaspi et al.\ 2001), B1957+20 (Stappers et al.\ 2003), J1509$-$5859, and
J1809$-$1917 (Sanwal et al.\ 2005) have revealed elongated structures (of lengths
$\sim 0.05$--0.5 pc) which look like ``trails'' behind moving pulsars.  An
exceptionally long ($\sim 1.6$ pc) trail was found behind PSR B1929+10 by
Wang, Li, \& Begelman (1993).
The origin of these trails has not been firmly established (see \S4).

An intriguing PWN structure has been recently reported by Caraveo et al.\
(2003; hereafter C03)
from 
an \xmm\ observation of
the middle-aged pulsar Geminga
($\tau=340$ kyr, $\dot{E}=3.3\times 10^{34}$ ergs s$^{-1}$).
They
found two $2'$-long 
tails behind the pulsar, with a luminosity of $\sim 10^{29}$
ergs s$^{-1}$ in the 0.3--5 keV band. They suggested that these are the tails
of a bow shock generated by pulsar's motion and predicted the head
of the bow-shock to be at an angular distance of $20''$--$30''$ ahead of the
pulsar.  In this paper, we present the results of our high-resolution
{\sl Chandra} observation of the Geminga PWN\footnote{ 
Preliminary results have been presented by Sanwal, Pavlov,
\& Zavlin (2004).} and confront them with the results of the
{\sl XMM-Newton} observation and theoretical models.

\begin{figure*}[ht]
\hskip 0.9cm
\includegraphics[height=7.2cm,angle=0]{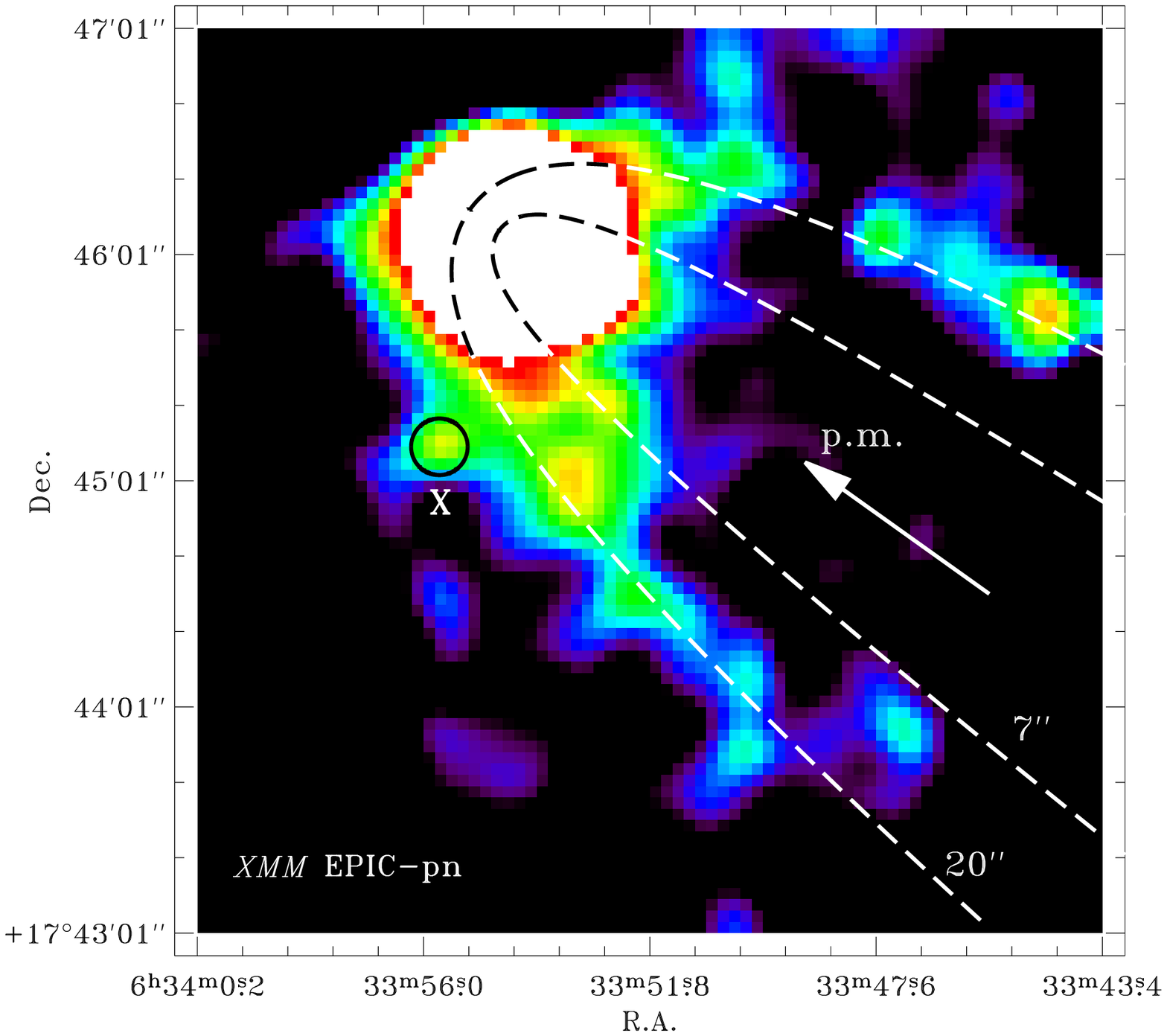}
\hskip 1.6cm
\includegraphics[height=7.2cm,angle=0]{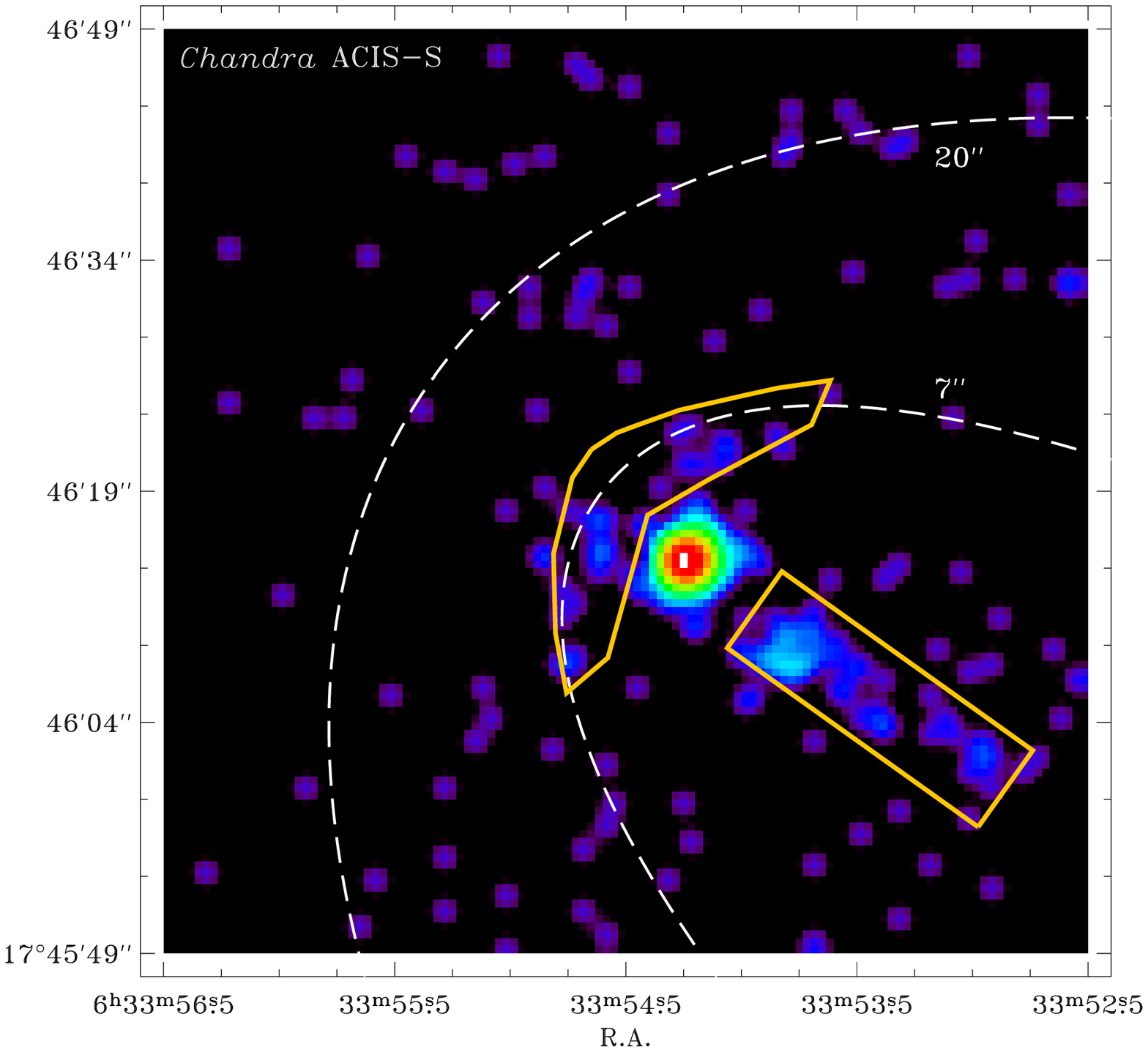}
\vskip 20truept
\caption{
{\em Left}:
\xmm\ EPIC-PN image of a $4'\times4'$
region around Geminga,
smoothed with a $15''$ FWHM Gaussian.
The dashed curves show two bow-shock models by Wilkin (1996),
with stand-off distances $20''$ and $7''$; the former fits the
shape of the two outer tails, while the latter corresponds to
the actually observed arc ahead of the moving pulsar (see the
right panel).
The arrow shows the  direction of pulsar's proper motion.
The black circle indicates the position of
the point source X resolved
in the {\sl Chandra} image
(see Fig.\ 2).
{\em Right}:
\chan\ ACIS-S3 image of a $1' \times 1'$
 region around
Geminga,
showing an extended emission at $5''$--$7''$
ahead of the pulsar (the arc)
and a
$20''$-long axial tail.}
\label{fig1}
\end{figure*}

\section{Observations and Data Reduction}

\chan\ observed Geminga on 2004 February 7 for 19.9~ks (18.8 ks good exposure
time).
The pulsar was positioned on ACIS-S3 chip
with a standard Y offset of $-0\farcm33$.  A 1/8 subarray mode (data taken
from a $1'\times 8'$ region of each CCD, with a frame time of
0.7 seconds) was used to mitigate the pileup in the pulsar image.  We used
the Chandra Interactive Analysis of Observations
(\texttt{CIAO}) software (ver.3.2.2; \texttt{CALDB} ver.3.1.0) 
for the analysis, starting from
the level 1 event files  to correct for charge transfer inefficiency. 
We applied the standard grade filtering 
and
removed events with 
energies $>8$ keV to reduce the background.
Since no events with $E<0.45$ keV were
telemetered from the S3 chip, we use the 0.5--8 keV
band for further analysis.
To account for nonuniform exposure and nonuniform CCD response near
the node boundaries, we applied the exposure map correction,
using
the 
\texttt{CIAO} script \texttt{merge\_all}.

We also reanalyzed the \xmm\ EPIC data taken on 2002 April 4--5 (103~ks
exposure, with EPIC-PN in Small Window mode and EPIC-MOS in Prime Full Window
mode).  We used \texttt{SAS} v.6.1 
to reprocess and analyze the data.  The effective
exposure times, after removing the periods of high background,
are about 68~ks for EPIC-PN and 82~ks for MOS1 and MOS2.  
A detailed analysis of the MOS data has been presented by C03.  We will supplement
those results with the analysis of the EPIC-PN data.

\section{Observational Results}
%%%%%%%%%%%%%%%%%%%%%%%%%%%%%%%%%%%%%%%%%%%%%%%%%%%%%
\subsection{XMM-Newton EPIC-PN Results}
%%%%%%%%%%%%
Smoothed MOS and PN images 
show two
elongated patchy structures,
stretched in the direction opposite
to the pulsar's proper motion.
We will call them {\em outer tails} to distinguish from
the short 
inner tail discovered with \chan\ (Sanwal et al.\ 2004; see \S3.2).  
Figure~1 (left panel) shows these structures,
seen up to about $3'$ from the pulsar
($\sim 0.17 d_{200}$ pc, where $d_{200}$ is the distance\footnote{
As mentioned by Kargaltsev et al.\ (2005), the Geminga's parallax
reported by  
Caraveo et al.\ (1996) is incorrect, and
the distance to Geminga
is currently unknown;
however, $d\sim 200$ pc seems to be
a reasonable estimate.} 
scaled to 200 pc).
To produce the 
image, we subtracted the intrinsic PN and particle
background using a closed-filter observation in 
Small Window mode 
taken on 
2002 December 30.
The raw data were binned
in $3''\times 3''$ pixels and smoothed with a Gaussian of FWHM = $15''$.
Subtracting the background from a $1'\times 1'$ box in the lower left
corner of the PN image, we found
the average image brightness of the tails 
(extraction box of $40'' \times 80''$ for each tail) about
$1\times 10^{-6}$ counts s$^{-1}$ arcsec$^{-2}$ in the 0.3--8 keV band.
Spectral analysis of the tail emission
is hindered by the 
strong, nonuniform background.  
Using
the absorbed power-law spectral model with the parameters estimated by C03
from the MOS data (photon index $\Gamma=1.6$, hydrogen column density
$N_{\rm H} = 1.1\times 10^{20}$ cm$^{-2}$), we obtain the surface brightness
(intensity) of the 
outer tails $I_{\rm outer} \sim 3\times10^{-18}$ ergs cm$^{-2}$ s$^{-1}$
arcsec$^{-2}$, and the total unabsorbed flux $F_{\rm outer}\sim 2\times10^{-14}$ ergs
cm$^{-2}$ s$^{-1}$.
  This gives the luminosity 
$L_{\rm outer} \sim 1 \times 10^{29} d_{200}^2\,{\rm ergs}\,{\rm s}^{-1}
\approx 3\times10^{-6} \dot{E}\, d_{200}^2$,
in agreement with 
the value reported by 
C03.

The tails in the PN image look similar to those in the MOS images,
reported by C03.
Their shape  
fits the same analytical bow-shock model by Wilkin (1996),
with 
an inclination angle $i\approx 90^\circ$ between the line of sight
and pulsar velocity
 and a stand-off
distance of about $20''$, 
hidden in the 
pulsar image broadened 
by the wide \xmm\ PSF. We note, however, that 
although such one-zone models can describe the FBS shape close
to its apex, they
are not expected to be applicable to PWN tails (see, e.g., B05).

Thus, the PN data show the properties of the outer tails
consistent with those derived by C03 from the MOS
data. 
However, the patchy structure of the tails hints that there may be
a substantial contribution from
faint background objects (such as the
point source `X' 
 resolved with {\sl Chandra}; see Fig.\ 2).
To separate such a contribution and prove that the tails are
not an artifact, they
should be observed with a better spatial resolution.

\begin{figure*}[ht]
\begin{center}
\includegraphics[height=8.9cm,angle=0]{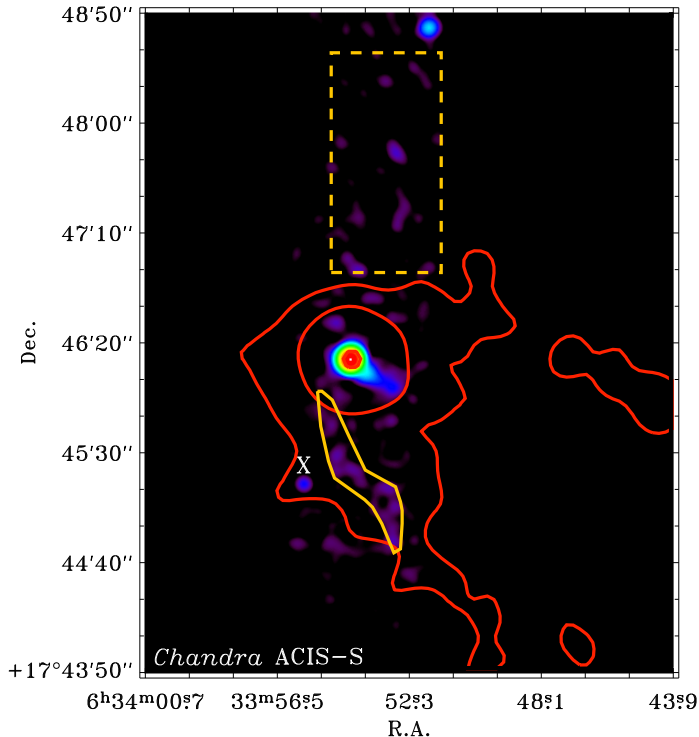}
\end{center}
\vskip 20truept
\caption{
ACIS-S3 image of
a $1'\times 5'$ region around Geminga, smoothed with a $7''$ FWHM Gaussian,
with overlaid PN brightness contours (red curves).
The polygon south of Geminga encloses the region of enhanced emission
apparently connecting the arc (not seen at this resolution) with
the southern outer tail. The dashed rectangle
shows the region used for the background evaluation.
The
`X' marks a point source
near the southern
outer tail.
}
\label{fig2}
\end{figure*}

\subsection{Chandra ACIS Results}

The excellent 
resolution of \chan\ provides a close-up view in the
vicinity of Geminga
(right panel of Fig.\ 1)
The image shows no
emission $20''$ ahead of the pulsar,
predicted by C03.
Instead, we see some
diffuse emission at a distance of $5''$--$7''$,
whose shape resembles an {\em arc}
extended perpendicular to the proper motion direction. 
The arc and the outer tails cannot be fitted with the same Wilkin's model.
The arc-like polygon of a 116.5 arcsec$^2$ area
contains 32 counts,
of which 10.1 counts are estimated to belong to the background
(0.087 counts arcsec$^{-2}$, as measured in a $49''\times 98''$ source-free
rectangle 
north of the pulsar, 
shown in Fig.\ 2).
This gives $21.9\pm 5.7$
background-subtracted counts.
The arc's spectrum ($\Gamma=1.2\pm 0.4$
at fixed $N_{\rm H}=1.1\times 10^{20}$ cm$^{-2}$, as obtained from a 
power-law fit
using the C-statistic) is substantially harder than
the pulsar's spectrum.
A structure of such size and spectrum cannot be ascribed to PSF tails of
the pulsar's image, which does not show any significant pileup
(the pulsar's count rate is only 0.07 counts/frame).
The average image brightness of 
$1.0\times 10^{-5}$ counts
s$^{-1}$ arcsec$^{-2}$
corresponds to an  
intensity  
$I_{\rm arc}\sim 0.9\times 10^{-16}$
ergs cm$^{-2}$ s$^{-1}$ arcsec$^{-2}$,
about 30 times brighter than the tails
seen with EPIC.  The X-ray luminosity of this structure is  
$L_{\rm arc}\sim 
5\times 10^{28}$ $d_{200}^2$ ergs s$^{-1}$.

The most striking feature in the ACIS image is a
$\sim 5''$-wide
{\em axial tail}, seen up to $25''$
($7.5\times 10^{16}d_{200}$ cm) from the pulsar
 in the direction opposite
to the pulsar's proper motion.
The tail is apparently detached from the pulsar by $5''$--$6''$;
its brightness is maximal at $\sim 8''$, and it fades with increasing
distance from the pulsar.
The average image brightness
in the $6''\times 12''$ box (see Fig.\ 1, right panel), which contains
$37.1\pm 6.6$
background-subtracted counts, is
$2.7\times10^{-5}$ counts s$^{-1}$ arcsec$^{-2}$.
Although the small number of counts
precludes
detailed spectral analysis,  
the tail's spectrum
can be described by a power-law model with
$\Gamma = 1.0\pm 0.2$,
somewhat harder than the spectrum of the
outer tails.
The intensity
of the tail
 (in the $6''\times 12''$ box),
$I_{\rm axial}
=3\times 10^{-16}$ ergs cm$^{-2}$ s$^{-1}$ arcsec$^{-2}$,
is about 2 orders of magnitude higher than that of the tails seen with EPIC.
Its luminosity, $L_{\rm axial}
= (1.3\pm 0.2)\times 10^{29}$ $d_{200}^2$ ergs
s$^{-1}$ (measured from $49.1\pm 7.3$ background-subtracted counts
in a $6''\times 20''$ box detached by $6''$ from
the pulsar),
 is close to the total luminosity of the putative
outer tails. 
An inspection of the EPIC PN image 
shows some patchy enhancements along the tail direction,
 at a level of $\lesssim 2\times 10^{-7}$ counts s$^{-1}$
arcsec$^{-2}$,
but they are indistinguishable
from background fluctuations. 

Heavily smoothed ACIS images show an enhancement south of the
pulsar which apparently connects the arc with the 
southern outer tail (an example is shown in Fig.\ 2).
The region of enhanced emission within the 
829 arcsec$^2$ polygon
includes 
105 counts. Subtracting 
71.9 counts of the scaled background,
we obtain 
$33.1\pm 10.8$ excess counts, which corresponds to 
an intensity
a factor of 6 higher than that
in the EPIC tails. 
The slope of the excess spectrum, $\Gamma = 1.0\pm 0.4$,
is apparently similar
to that of the axial tail.
Thus, although the $3\,\sigma$ enhancement can hardly be considered
as a firm detection, 
its position and shape
support the reality of both this structure and the southern outer
tail. 
No enhancement is seen at the site of the
northern outer tail; however, the northern tail looks less significant
and more patchy in the EPIC images, and only a small part of it was
imaged with ACIS.

\section{Discussion}

If the tails in the EPIC images are real, the Geminga PWN is truly unique: a
bow-shock-like structure with long 
outer tails {\em and} a short axial tail
behind the pulsar have never been seen before in X-rays. 
Before discussing possible interpretations of the observed structures, we note
that the proper motion of Geminga, 0\farcs17 yr$^{-1}$, implies a pulsar speed
$v = 160\,\tilde{d}$ km s$^{-1}$, where $\tilde{d} = d_{200}/\sin i$.  For a
reasonable distance,
it exceeds a typical
sound speed in the interstellar medium (ISM),
$c_s = 15\, (\mu/0.6)^{-1/2} T_4^{1/2}$ km s$^{-1}$,
where $\mu$ and $T = 10^4 T_4$ K are the molecular weight and temperature.
Assuming that the speed of a possible ISM flow at the location of Geminga is
much lower than $v$, the ram pressure due to the pulsar motion in the ISM is
$p_{\rm ram} = 4.3\times 10^{-10} n\, \tilde{d}^{2}$ ergs cm$^{-3}$,
where $n$ is the ISM density in atomic mass units per cm$^3$,
  This gives an estimate
$R_{\rm h} = 1.4\times 10^{16} n^{-1/2} \tilde{d}^{-1}$ cm for the
stand-off distance of the TS head, which translates into the projected angular
distance ${\cal R}_{\rm h} = 4\farcs8\, n^{-1/2} \tilde{d}^{-2}$.  Thus, one
can expect that Geminga is accompanied by a bow-shock PWN, with a characteristic
size comparable to the sizes of the structures observed with \chan.
We will discuss possible interpretations of the whole PWN, starting each from
an assumption on the nature of the axial tail, the brightest 
feature of the PWN.

\subsection{The axial tail is a shocked pulsar wind?}

The axial tail could be interpreted as synchrotron emission from the 
shocked
PW collimated by the ram pressure.  According to the simulations 
by B05,
who assumed
an isotropic PW, the TS has a bullet-like shape.  For large Mach numbers,
${\mathcal M} = v/c_s$, and small values of the
 magnetization parameter $\sigma$ of the pre-shock
PW (see Kennel \& Coroniti 1984), 
the bullet's cylindrical radius is $r_{\rm TS} \sim R_{\rm h}$, and the
distance of its back surface from the pulsar is $R_{\rm b} \sim 6 R_{\rm h}$.
The shocked PW outside the TS is confined inside the CD surface which has a
cylindrical shape behind the TS, with a radius $r_{\rm CD} \sim 4 R_{\rm h}$.
The collimated PW flows with subrelativistic velocities: 0.1--0.3~$c$ in the
inner channel, $r\lesssim r_{\rm TS}$, and up to
 0.8--0.9~$c$ in the outer channel,
$r_{\rm TS} \lesssim r \lesssim r_{\rm CD}$
(see Figs.\ 1--3 and \S3.3 in B05).

First, one can speculate that the axial tail is the CD-confined cylindrical
tube behind the TS, which implies 
a CD radius of 
$\sim 3''$, 
${\cal R}_{\rm h} \sim 
0\farcs7\sin i$, and ${\cal R}_b \sim 5''\sin i$.  In this interpretation, one
should expect brightest emission from the shocked PW at $\lesssim 1''$ ahead of
the pulsar, hidden within the pulsar image. The actually observed emission
$\sim 5''$--$7''$ ahead of the pulsar (the arc) is not explained by this model.
Being well outside the CD, the two 
outer tails 
cannot be associated with a shocked PW.  One might speculate that they
are produced by the shocked ISM heated to X-ray temperatures,
but
the pulsar's speed is too low to support this
speculation, and thermal models with reasonable temperatures do not fit the
tails spectrum.  Overall, 
given the problems with explaining the observed PWN
structure, 
this interpretation of the axial tail is hardly viable.

Second, one could assume that the axial tail is associated with the shocked PW
``sheath'' immediately outside the bullet-like TS, 
similar to the interpretations of the
tail behind PSR B1757--24 by Gvaramadze (2002) and the ``tongue'' of the X-ray
Mouse PWN by Gaensler et al.\ (2004).  The two 
outer tails might be associated
with emission from a shell immediately inside the CD surface, where the magnetic
field is compressed (see B05) and hence the synchrotron emissivity could be
enhanced\footnote{Note, however, that the simulation of 
intensity by
B05 do not show bright shell-like structures.}.  In this interpretation, the 
outer 
tails should be parallel to the direction of pulsar motion, which is crudely in
agreement with observations, but their distance from the PWN axis,
$\sim 60''$--$70''$, 
implies a bullet diameter $\sim 30''$--$35''$,
much larger than the observed width of the axial tail.
This interpretation also implies ${\cal R}_{\rm h} \sim 16''\sin i$,
which is larger than the distance to the observed arc unless the inclination
angle is small, $\sin i \lesssim 0.3$.
On the other hand, the 
width 
of the axial tail suggests
${\cal R}_{\rm h} \sim 2\farcs5 \sin i$, 
smaller than the distance to the arc in the {\sl Chandra} image.  Moreover,
the arc is, on average, a factor of 4 dimmer than the tail, and its extent
perpendicular to the pulsar's proper motion is a factor of 
4 larger
than the tail's width.  Therefore, we are forced to assume that the TS head is
unresolved from the pulsar, while the arc might be a CD head at a distance
$\sim 2.8 R_{\rm h}/\sin i$ from the pulsar, larger than $\sim 1.3 R_{\rm h}$ in the
B05 simulations.  The corresponding ratio  $r_{\rm CD}/R_{\rm h} \sim 
28/\sin i$ is much larger than $\sim$4 in the B05 simulations.
The simulations also do not 
explain the apparent detachment of the axial tail from the pulsar.
 These discrepancies could
be caused by {\em anisotropy of the pulsar outflow}, neglected by B05.
For instance, if the outflow is mostly
equatorial (assuming the pulsar's spin axis aligned with its space velocity),
the flattened TS head should be closer to the pulsar than in the isotropic case.

We should also mention that this interpretation implies a 
rather large tail's length,
$l_{\rm tail}\sim v_{\rm flow} \tau_{\rm syn}$, where $v_{\rm flow}$ is the flow
velocity and $\tau_{\rm syn}
=5.1\times 10^8 \gamma^{-1} B^{-2}$ s
is the synchrotron cooling time. 
The magnetic field just downstream of the TS back boundary
can be estimated, for $\sigma\ll 1$, as
\begin{equation}
B_{\rm b} \simeq 3 \left(\frac{\dot{E}\sigma}{R_{\rm b}^2 c}\right)^{1/2}
\simeq 70\, \frac{15''}{{\cal R}_{\rm b}}\, \frac{\sigma^{1/2}}{\tilde{d}}\,\mu{\rm G}
\end{equation}
(cf.\ Kennel \& Coroniti 1984).
The synchrotron photons with maximum observed
energies $E\approx 8$ keV are emitted by electrons with a Lorentz factor 
\begin{eqnarray}
\gamma & \sim & 1.3\times 10^8 \left(\frac{E}{8\,{\rm keV}}\right)^{1/2}
\left(\frac{100\,\mu{\rm G}}{B}\right)^{1/2} \nonumber \\
 & \sim &
1.5\times 10^8 \left(\frac{E}{8\,{\rm keV}}\right)^{1/2}
\left(\frac{{\cal R}_{\rm b}}{15''}\right)^{1/2} \frac{\tilde{d}^{1/2}}{\sigma^{1/4}}\,.
\end{eqnarray}
This gives 
\begin{equation}
l_{\rm tail} \sim 4\times 10^{18}\frac{v_{\rm flow}}{0.2\,c} 
\left(\frac{8\,{\rm keV}}{E}\right)^{1/2} \left(\frac{{\cal R}_{\rm b}}{15''}\right)^{3/2} \frac{\tilde{d}^{3/2}}{\sigma^{3/4}}\, {\rm cm},
\end{equation}
which exceeds the observed length by two orders
of magnitude even at lowest $v_{\rm flow}\sim 0.1\,c$ in the
inner channel behind the TS, as found by B05. This discrepancy could
be explained assuming that the surface brightness of the tail becomes
too low at large distances from the pulsar (e.g., because of a
decreasing magnetic field) to be seen in these images.

Thus, the interpretation 
that the axial tail is 
the freshly shocked PW immediately outside the TS while the 
outer tails mark the CD surface is not quantitatively consistent
with the available simulations.
However, such an interpretation  
cannot be ruled out because those simulations do not take into
account the intrinsic anisotropy of the PW.

\subsection{The axial tail is a pulsar jet?}

Another explanation of the axial tail, which we consider more plausible,
is that it is a jet emanating from
the pulsar magnetosphere along the spin axis aligned with the pulsar's motion.
The fact that only one jet is seen is not uncommon (PSR B1706$-$44 is a vivid
example; Ng \& Romani 2004), and it can be explained by Doppler boosting
(the approaching jet is brighter than the receding one) and/or by intrinsic
anisotropy of the polar outflows, or it may be caused by destruction of the
forward jet by the ISM ram pressure.  The Geminga's axial tail resembles
the southeast jet (``inner counterjet'' in Pavlov et al.\ 2003) of the Vela
PWN, which is about twice brighter than the northwest jet in the direction of
pulsar's motion. The Vela's southeast jet is also somewhat detached from
the pulsar (perhaps because the polar outflow becomes visible only beyond
a shock). Its projected length,
$5\times 10^{16}d_{300}$
cm, is close to that of the Geminga's axial tail,
the spectra of both structures are very hard,
and the ratios of their X-ray luminosities 
to the pulsar spindown powers are not very different: 
$L_X/\dot{E} \sim 3.6\times 10^{-6} d_{200}^2$ for the Geminga's tail and
$\sim 0.6\times 10^{-6}d_{300}^2$ for the Vela's southeast jet.

If the jet is confined by its own magnetic field, a lower limit on the field
can be estimated from the requirement that the electron Larmor radius is smaller
than the jet radius, $r_{\rm jet}\sim 0.75\times 10^{16} d_{200}$ cm, which
gives 
\begin{equation}
B > 90 \left(\frac{E_{\rm M}}{8\,{\rm keV}}\right)^{1/3} d_{200}^{-2/3}\,\, \mu{\rm G},
\end{equation}
where
$E_{\rm M}$ is the maximum energy of the X-ray power-law spectrum (cf.\ Pavlov
et al.\ 2003).  It corresponds to the energy injection rate 
\begin{eqnarray}
W & = & \frac{B^2}{8\pi} (1+k) v_{\rm jet} \pi r_{\rm jet}^2 \nonumber \\
 & > & 7.8\times 10^{32} (1+k)\, \frac{\beta}{0.5} \left(\frac{E_{\rm M}}{8\,{\rm keV}}\right)^{2/3} d_{200}^{2/3}\,\, \frac{\rm ergs}{\rm s}\, ,
\end{eqnarray}
where $v_{\rm jet} =\beta c$ is the
bulk flow velocity in the jet, and $k$ is the ratio of particle and magnetic
energy densities.  Even this lower limit on $\dot{W}$ is a substantial fraction
of the spindown power for a mildly relativistic $v_{\rm jet}$ expected 
(e.g., $\dot{W}\sim 0.05\dot{E}$ for $k=1$ and $\beta =0.5$), which
means that the magnetic field cannot strongly exceed the above lower limit.
At such magnetic fields one would expect a jet length
\begin{eqnarray}
l_{\rm jet} & \sim & v_{\rm jet}\tau_{\rm syn} \nonumber \\
 & \sim & 2\times 10^{18} \frac{\beta}{0.5} \left(\frac{100\,\mu{\rm G}}{B}\right)^{3/2} \left(\frac{8\,{\rm keV}}{E_{\rm M}}\right)^{1/2} {\rm cm},
\end{eqnarray}
much larger than the observed $\sim 7\times 10^{16} \tilde{d}$ cm.
Therefore, we have to assume that the jet is destroyed or becomes
uncollimated well before it loses its internal energy to radiation.

If the axial tail is a pulsar jet, the 
outer tails could mark an equatorial
outflow bent by the ram pressure. We are unaware of theoretical PWN models 
which include both the PW anisotropy and the ram pressure effects.
We, however, expect that the equatorial PW component (which would produce a torus
beyond a TS ring around a slowly-moving pulsar) would form a relatively thin
shell between the TS and CD surface behind the pulsar, filled by a relativistic
plasma with a subrelativistic bulk flow velocity.  If the 
outer tails turn
out to be an artifact, we would suggest that most of the PW flows out of the
magnetosphere along the spin axis while the equatorial PW component is
unusually weak in Geminga (perhaps because of a small angle between the magnetic
and spin axes).  The arc ahead of the pulsar could be a head of the bent
equatorial outflow or remnants of a forward jet crushed be the ISM ram pressure.

To summarize, the \chan\ observation has conclusively shown the presence of
PWN elements around the Geminga pulsar.  With the sparse statistics of
the \chan\ data and the low spatial resolution of the \xmm\ images, we cannot
establish the nature of the extended emission unambiguously.  However, it
seems certain that the observed PWN structure implies that the Geminga's PW
is intrinsically anisotropic.  Much deeper \chan\ observations 
and modeling of magnetized anisotropic winds from fast-moving
pulsars are needed to clarify the nature of this intriguing PWN\footnote{
After our work was mostly completed, we became aware of the paper by
De Luca et al.\ (astro-ph/0511185) who reported on the same {\sl Chandra}
data. They found virtually the same properties of the axial tail
but  failed to notice the arc and the enhancement connecting the
arc with the southern outer tail.}.
 
\acknowledgements{}
We acknowledge useful discussions with Niccolo Bucciantini and thank
Roger Romani, the referee, whose remarks allowed us to present
the results more clearly.
Support for this work was provided by the NASA 
through Chandra Award 
Number 
GO4-5083X issued by the Chandra X-ray Observatory Center,
which is operated by the Smithsonian Astrophysical Observatory for and 
on behalf of the NASA under contract NAS8-03060.
The work of G.G.P. and D.S. was also partially supported by NASA grant NAG5-10865.
The work of V.E.Z. is supported by a NASA Fellowship Award
at NASA MSFC.

\end{document}